\documentclass{article}
 \pdfoutput=1
\usepackage{PRIMEarxiv}

\usepackage[utf8]{inputenc} 
\usepackage[T1]{fontenc}    
\usepackage{hyperref}       
\usepackage{comment}
\usepackage{url}    
\usepackage{longtable}
\usepackage{booktabs}       
\usepackage{amsfonts}       
\usepackage{nicefrac}       
\usepackage{microtype}      
\usepackage{lipsum}
\usepackage{fancyhdr}       
\usepackage{graphicx}       
\graphicspath{{media/}}     
\usepackage{amssymb,amsmath}
\usepackage{amsthm}
\newtheorem{lemma}{Lemma}
\numberwithin{equation}{section}
\theoremstyle{definition}
\newtheorem{thm}{Theorem}[section]

\newtheorem{exmp}{Example}[section]
\title{$D$-optimal Approximate Design for Binary Regression and Quantal Response in Toxicology Studies
}

\author{
  Elvis Han Cui\\
  Department of Biostatistics, UCLA \\
  \texttt{elviscuihan@g.ucla.edu} \\
}

\begin{document}
\maketitle

\begin{abstract}
We provide a systematic treatment of $D$-optimal design for binary regression and quantal response models in toxicology studies. For the two-parameter case, we provide an analytical equation (WC equation) for computing the $D$-optimal design quickly and when analytical solution is not available, we apply particle swarm optimization to solve for the $D$-optimal design. Examples with various link functions are given as well as the sensitivity functions. We extend the two-parameter case to three-parameter case by providing a neat formula for the determinant of the information matrix. We also suggest practitioners to work with the neat formula to derive optimal designs for three-parameter binary regression models.
\end{abstract}

\keywords{$D$-optimality \and Quantal response \and Binary regression \and WC equation}

\section{Preliminaries}\label{sec:concepts_opt_design}
In this section, we first briefly discuss quantal response in toxicology studies and then introduce the motivation and the basic concepts of optimal approximate design and review binary regression for toxicology studies.

\subsection{Quantal Response in Toxicology Studies}
\begin{figure}[h]
\centering
\includegraphics[width=12cm]{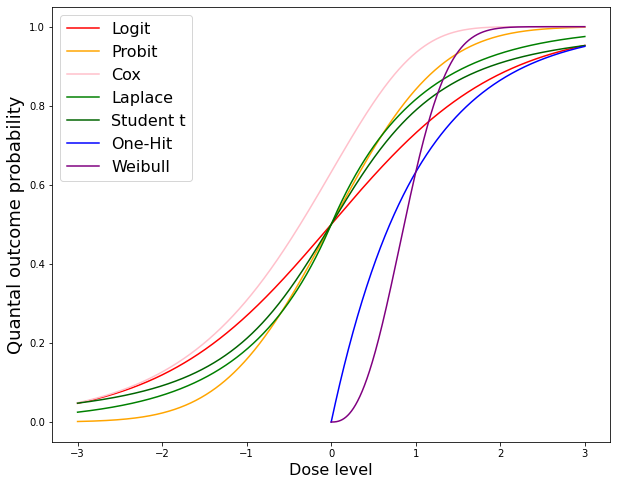}
\caption{Dose response curves for different link functions.}
\label{fig:dose_resp_link}
\end{figure}

The outcome of interest in many toxicological studies is qualitative in nature. In a vasr majority of these experiments, the outcome is quantal and binary \cite{razzaghi2022statistical}, to name a few: beetle mortality and embryogenic anthers in toxicology studies \cite{dobson2018introduction}; tumor progression status in cancer studies \cite{antonios2017immunosuppressive}; low-density lipo-protein (LDL) cholesterol levels (desirable vs undesirable) \cite{collins2022model}. Further, in developmental toxicity studies, pregnant female animals are being exposed to a teratogen during a specific time of pregnancy \cite{razzaghi2022statistical}. To estimate the potential risk from a defined source of hazard (e.g., exposure to a teratogen) quantitatively and qualitatively, we usually starts with fitting a dose-response curve to the data. For quantal and binary outcome, the range of the curve is within 0 to 1. In Figure~\ref{fig:dose_resp_link}, we demonstrate the dose-response curve using 7 different potential ``link functions'' (for details, see \ref{sec:binary_reg}), the $x$-axis refers to potential dose level (potential means the range can be shifted and re-scaled) and the $y$-axis refers to the binary outcome probability. As Figure~\ref{fig:dose_resp_link} has shown, when the dose increases, the probability of the quantal / binary outcome is monotonically increasing. 

We are particularly interested estimating the curve as accurate as possible so that the downstream risk assessment \cite{razzaghi2022statistical} can proceed smoothly. Hence, it is desirable for us to estimate the curve as accurate as possible and simultaneously reduce the required total number of doses \cite{wong1996designing, zhu1998multiple, zhu2000multiple, baek2006bayesian}. To achieve this goal, we apply the theory of optimal design and we give a brief introduction in section \ref{sec:basic_opt_design}.

\subsection{Motivation and Basic Concepts of Optimal Design}\label{sec:basic_opt_design}
In a dose–response experiment, decisions regarding the dose range, the number of doses, the dose levels, and the number of experimental units at each dose are sometimes made predicated on nebulous criteria. These are design issues that can potentially have a substantial impact on the quality of the statistical inference at the end of the study, yet they are decided in some cases on an ad-hoc basis. Frequently, an equal number of experimental units are assigned at each dose. When the doses are equally spaced, these are called uniform designs in the statistical literature and while they are appealing and intuitive, it has been shown that they can be inefficient, depending on the goal of the study and the underlying model assumed. For example, \cite{wong1996designing} showed that performance of such designs can depend sensitively on the choice of the number of doses in a uniform design, the model, and the optimality criteria. Therefore, each aspect in the design of the study must be carefully considered to realize maximum accuracy in the information. Such attention to detail will enhance reproducibility, thus addressing a current issue in animal experimentation \cite{giles2006animal} and reducing the overall cost of experiments. More specifically, if the current cost for producing a new drug is 10 dollars per dose, then using optimal design theory, one is able to reduce the cost to 5 dollars per dose.

To optimally design an experiment, model assumptions are required to work out the mathematical and statistical details. Invariably, the goal is formulated as an objective function defined on the user-specified dose range (or design interval) that depends on the statistical model and the design. The optimization of the criterion can then be performed among a specific class of designs, for example, among all designs with five doses, or among all designs on a given dose interval. The resulting optimal design is therefore model-based and, as a consequence, can be highly model-dependent, suggesting that choice of a statistical model for the dose–response study is also important.

Optimal approximate designs in clinical trials can help investigators achieve higher quality results for the given resource constraints \cite{schwaab2006new, jozwiak2013podse, sverdlov2020optimal, zhou2021optimal}. The creation of this field can be traced back to \cite{smith1918standard}. From 1950s to 1980s, the field of approximate design has witnessed a booming development \cite{federov1972theory, kiefer1974general, pazman1986foundations, atkinson2007optimum, silvey2013optimal} and we give a brief introduction below.

Consider a linear model $\mathbb{E}(y)=\mu=\beta^Tf(x)$ where $\mu$ is the expectation of $y$. A $k$-point design $\xi$ is a $2\times k$ matrix of the following form
\begin{align*}
    \xi&=\left(\begin{matrix}
    x_1&x_2&\cdots& x_{k-1}&x_k\\
    p_1&p_2&\cdots& p_{k-1}&p_k
    \end{matrix}\right)
\end{align*}
where $x_i$'s are called design points and $p_i$'s are non-negative weights that sum to 1. In practice, we usually have a total of $n$ observations and $np_i$ is not an integer in general. Hence, we choose the closest integer of $np_i$ and assign the corresponding dose $x_i$ to these individuals. For this reason, $\xi$ is referred as approximate design in literature. The information matrix associated with design $\xi$ is
$$\mathfrak{M}(\xi)=\int_\mathcal{X}f(x)f(x)^T\xi(dx)=\sum_{i=1}^kp_if(x_i)f(x_i)^T$$
where $\xi(dx)$ is the measure induced by $p_i$'s. The optimal design seeks to find a $\xi^*$ that minimizes $\phi(\mathfrak{M}(\xi))$ where $\phi(\cdot)$ is a real-valued function. Commons choices of $\phi(\cdot)$ and their terminologies are given in the following table

\begin{longtable}{p{0.15\linewidth}lr}
\caption{List of Common Choices of $\phi(\cdot)$}\\
\toprule
\textbf{Optimality}\textsuperscript{a} & \textbf{Choice of} $\phi$ 
 & \textbf{Remarks} \\ 
\midrule 
A & $\text{Tr}(\mathfrak{M}^{-1})\textsuperscript{b}$ & Sum of variances  \\
c & $\text{Var}(g(\widehat{\beta}))\textsuperscript{c}$ & Variance of $g(\widehat{\beta})$\\
D & $\log\det\mathfrak{M}^{-1}$ & Log-volume of the ellipse \\
E & $\min\lambda_i(\mathfrak{M})=\lambda_{min}\textsuperscript{d}$ & Length of minor axis\\
G & $\max_{i}(\mathfrak{M}^{-1})_{ii}$ & Maximum of $\text{Var}(\widehat{\beta})$\\
I & $\int_{\mathcal{X}}f(x)^T\mathfrak{M}^{-1}f(x)\mu(dx)$ & Integrated variance \\
\bottomrule
\label{tab:optimality}
\end{longtable}
\vspace{-5mm}
\begin{minipage}{\linewidth}
\textsuperscript{a} For example, the first line reads $A$-optimality.

\textsuperscript{b} $\text{Tr}$ refers to the trace function of a matrix.

\textsuperscript{c} $g$ is a function of $\beta$, $\widehat{\beta}$ is the MLE of $\beta$ and the variance of $g(\widehat{\beta})$ can be derived using Delta method, i.e., $\widehat{\text{Var}}(g(\widehat{\beta}))= \nabla g(\widehat{\beta})^T\mathfrak{M}^{-1}\nabla g(\widehat{\beta})$ and $\nabla$ refers to the gradient operator.

\textsuperscript{d} $\lambda$'s refer to eigenvalues of $\mathfrak{M}$.
\end{minipage}
\\\\
Further, if we extend the linear model framework to generalized linear models, then the information matrix depends on parameters (see section \ref{sec:two_para_reg}). In this case, one usually plug-in plausible parameter values and then calculate the optimal design $\xi^*$. We call it $\phi$-optimal design, where $\phi$ refers to $D$-, $A$-, $c$-, $E$-, etc. In practice, researchers would like to consider different criteria simultaneously, leading to the so-called compound criteria. For a comprehensive review of different optimality criteria, see Chapter 10 of \cite{atkinson2007optimum} or the review paper by \cite{fedorov2010optimal}.

To verify that the resulting design is globally optimal, that is, optimal among all possible designs, one needs to apply the equivalence theorem \cite{kiefer1974general} and plot the sensitivity functions to check. Different optimality criteria corresponds to different types of equivalence theorems and sensitivity functions, hence for brevity, we only state the equivalence theorem for $D$-optimal design in section~\ref{sec:binary_reg}. In addition, \cite{chen2022particle} provides a comprehensive review on the application of PSO in optimal approximate design. 

\subsection{Basic Concepts of Binary Regression}\label{sec:binary_reg}
In the following, we first give a review on binary regression and then provide the equivalence theorem for $D$-optimal design. Assume that for $i=1,\cdots,n$, the response $y_i$ is a binary outcome with covariate $x_i\in\mathbb{R}^d$. The $y_i$'s are independently distributed and the density is
$$p(y|x,\pi)=p(y|\eta(x,\beta))=\exp\left(y\ln\frac{\pi}{1-\pi}+\ln(1-\pi)\right)$$
where $\beta$ is a $p$-dimensional parameter of interest and $\pi=F(\eta)=\int_{-\infty}^\eta f(s)ds$. Let $\mathcal{X}$ be the design space and the design $\xi$ be
\begin{align*}
    \xi&=\left(\begin{matrix}
    x_1&x_2&\cdots& x_{n-1}&x_n\\
    p_1&p_2&\cdots& p_{n-1}&p_n
    \end{matrix}\right)
\end{align*}
where for all $i$, $x_i\in\mathcal{X}$, $p_i\ge 0$ and $\sum_{i=1}^np_i=1$.
Let $g(x,\beta)=\mathbb{E}\left(-\frac{\partial^2 \log p(y|x,\pi)}{\partial\beta\partial\beta^T}\right)$ be the Fisher information associated with a single point $x$, then 
    $$g(x,\beta)=\omega xx^T$$
    where $\omega=F'(\eta)^2/(\pi(1-\pi))=f(\eta)^2/(\pi(1-\pi))$. The information matrix associated with the design $\xi$ is
\begin{align}
    \mathfrak{M}(\xi)&=\int_{\mathcal{X}}g(x,\beta)\xi(dx)\nonumber\\
    &=\sum_{i=1}^np_i\omega_ix_ix_i^T,\label{eq:fisher_info_binary_reg}
\end{align}
where
$$\omega_i=\frac{f(\eta_i)^2}{(\pi_i(1-\pi_i))},\ \sum_{i=1}^np_i=1,\ p_i\ge0.$$
Here are some examples of commonly used models in practice.
\begin{itemize}
    \item (Logit) The most famous model is the logistic regression with logit link.
    \begin{align*}
    f(\eta_i)&=\frac{\exp(\eta_i)}{(1+\exp(\eta_i))^2},\\
    \eta_i&=\ln\frac{\pi_i}{1-\pi_i},\\
    \pi_i&=\frac{\exp(\eta_i)}{1+\exp(\eta_i)},\\
    \mathfrak{M}(\xi)&=\sum_{i=1}^n\frac{p_i\exp(\eta_i)}{(1+\exp(\eta_i))^2}x_ix_i^T.
    \end{align*}
    
    \item (Probit) Prior to the presense of logit link, one uses the probit link.
    \begin{align*}
        f(\eta_i)&=\frac{1}{\sqrt{2\pi}}\exp\left(-\frac{1}{2}\eta_i^2\right),\\
        \eta_i&=\Phi^{-1}(\pi_i),\\
        \pi_i&=\Phi(\eta_i),\\
        \mathfrak{M}(\xi)&=\sum_{i=1}^n\frac{p_i\exp(-\eta_i^2)}{2\pi\Phi(-\eta_i)\Phi(\eta_i)}x_ix_i^T.
    \end{align*}
    where $\Phi(\cdot)$ is the cumulative function of standard normal.
        \item (Laplace) If we want the rate of decay is faster than student $t$ but slower than probit, then Laplace density is an alternative:
    \begin{align*}
        f(\eta_i)&=\frac{1}{2}\exp\left(-|\eta_i|\right),\\
        \eta_i&=F^{-1}(\pi_i),\\
        \pi_i&=F(\eta_i)=\frac{1}{2}+\frac{1}{2}\text{sgn}(\eta_i)\left(1-\exp(-|\eta_i|)\right),\\
        \mathfrak{M}(\xi)&=\sum_{i=1}^n\frac{p_i\exp(-2|\eta_i|)}{4F(\eta_i)S(\eta_i)}x_ix_i^T
    \end{align*}
    where $\text{sgn}(\cdot)$ is the sign function and $S(\cdot)=1-F(\cdot)$.
    \item  (Cox regression) The associated distribution is called the {Gumbel extreme value distribution}, and the link function is called the {complementary log-log}:
    \begin{align*}
        f(\eta_i)&=\exp\left(\eta_i-\exp(\eta_i)\right),\\
        \eta_i&=\ln\left(-\ln(1-\pi_i)\right),\\
        \pi_i&=1-\exp(-\exp(\eta_i)),\\
        \mathfrak{M}(\xi)&=\sum_{i=1}^n\frac{p_i\exp(2\eta_i-2\exp(\eta_i))}{\exp(-\exp(\eta_i)-\exp(-2\exp(\eta_i)))}x_ix_i^T.
    \end{align*}
    \item (Student t) The Student t distribution with degrees of freedom $k$ is useful in hypothesis testing and its density is symmetric with respect to 0:
    \begin{align*}
    f(\eta_i)&=\frac{\Gamma(\frac{k+1}{2})}{\sqrt{k\pi}\Gamma(\frac{k}{2})}\left(1+\frac{\eta_i^2}{k}\right)^{-\frac{k+1}{2}}\\ \eta_i&=F^{-1}(\pi_i)\\
    \pi_i&=F(\eta_i)=1-\frac{1}{2}\mathbf{I}_{x(\eta_i)}\left(\frac{k}{2},\frac{1}{2}\right)
    \end{align*}
    where $\mathbf{I}$ is the regularized incomplete Beta function and $x(s)=\frac{k}{s^2+k}$.
    \end{itemize}
A $D$-optimal design seeks to find a design $\xi^*$ such that $\det\mathfrak{M}(\xi)$ is maximized. It is well known that if we know the $D$-optimal design for a $k$-parameter model is supported at $k$-points, then all points are equally weighted \cite{atkinson2007optimum, wong2021lecnotes}.

\begin{lemma}[Property of $D$-optimal design]\label{thm:property_d_optim} If we know in advance that a $D$-optimal design for binary regression (Formula~\ref{eq:fisher_info_binary_reg}) has $k$ support points, then all design points have design weight $\frac{1}{k}$.
\end{lemma}
\begin{proof}
Suppose $n=k$ then we can write the determinant of formula~\ref{eq:fisher_info_binary_reg} as
\begin{align*}
    \det\mathfrak{M}(\xi)&=\det\left(\sum_{i=1}^kp_i\omega_ix_ix_i^T\right)\\
    &=\det\left(\left(\begin{matrix}\sqrt{\omega_1}x_1&\cdots&\sqrt{\omega_k}x_k\end{matrix}\right)
    \left(\begin{matrix}
    p_1&0&\cdots&0\\
    0&p_2&\cdots&0\\
    \vdots&\vdots&\ddots&\vdots\\
    0&0&\cdots&p_k
    \end{matrix}\right)\left(\begin{matrix}\sqrt{\omega_1}x_1\\\cdots\\\sqrt{\omega_k}x_k\end{matrix}\right)\right)\\
    &=\det\left(\sum_{i=1}^k\omega_ix_ix_i^T\right)\det\left(\begin{matrix}
    p_1&0&\cdots&0\\
    0&p_2&\cdots&0\\
    \vdots&\vdots&\ddots&\vdots\\
    0&0&\cdots&p_k
    \end{matrix}\right)\\
    &=\det\left(\sum_{i=1}^k\omega_ix_ix_i^T\right) \left(\prod_{i=1}^kp_i\right)\\
    &\le \det\left(\sum_{i=1}^k\omega_ix_ix_i^T\right)\left(\frac{\sum_{i=1}^kp_i}{k}\right)^k\\
    &=\det\left(\sum_{i=1}^k\omega_ix_ix_i^T\right)\left(\frac{1}{k}\right)^k
\end{align*}
where the inequality is due to the AM–GM inequality \cite{hardy1952inequalities}. The ``='' is attained if and only if $p_1=p_2=\cdots=p_k=\frac{1}{k}$. Hence, we conclude that all design points have equal weights.
\end{proof}
Finally, to check a design $\xi^*$ is whether globally optimal or not (i.e., optimal among all possible designs), we use the following theorem and plot the sensitivity function $\psi$.
\begin{thm}[Equivalence theorem \cite{atkinson2007optimum, wong2021lecnotes}]\label{thm:equivalence} Let $\mathfrak{M}(\xi)$ be the information matrix associated with design $\xi$,
then the following are equivalent ($\dim(\mathcal{X})=d$),
\begin{enumerate}
    \item The design $\xi^*$ is $D$-optimal, i.e., $\xi^*=\arg\min_{\xi}\det\mathfrak{M}(\xi)$.
    \item The inequality $\psi(\mathbf{x},\beta)=w(\mathbf{x})(\mathbf{x}^T\mathfrak{M}(\xi^*)^{-1}\mathbf{x})-d\le 0$ holds for all $\mathbf{x}\in\mathcal{X}\subset\mathbb{R}^d$ where $w(\mathbf{x})$ is a weight depending on the link $\eta(\mathbf{x})$ and $\psi$ is called the sensitivity function.
\end{enumerate}
\end{thm}

The equivalence theorem says that if a design $\xi^*$ is $D$-optimal, the the sensitivity function $\psi(\mathbf{x},\beta)$ is less or equal to 0 within the design space $\mathcal{X}$. Further, the sensitivity function attains 0 at the design points. 
Some preliminary work on $D$-optimal design for binary regression is given in \cite{king2000minimax, baek2006bayesian, haines2007d, atkinson2007optimum, kabera2012note}. However, there lack a detailed and unified framework for binary regression with different types of link functions under $D$-optimality. Hence, we provide a systematic treatment in the next two sections.

\section{Two-parameter Binary Regression}\label{sec:two_para_reg}
In this section, we always assume $k=2$ so that the resulting design always has $2$ equally support points. Then by formula~\ref{eq:fisher_info_binary_reg}, we have
\begin{align}
\det\mathfrak{M}(\xi)&=\frac{1}{4}\det\left(\begin{matrix}
    \omega_1+\omega_2&\omega_1x_1+\omega_2x_2\\
    \omega_2x_2+\omega_1x_1&\omega_1x_1^2+\omega_2x_2^2
    \end{matrix}\right)\nonumber\\
    &\propto\left(\omega_1+\omega_2\right)\left(\omega_1x_1^2+\omega_2x_2^2\right)-\left(\omega_1x_1+\omega_2x_2\right)^2\nonumber\\
    &=\left[\omega_1^2x_1^2+\omega_1\omega_2(x_1^2+x_2^2)+\omega_2^2x_2^2\right]-\nonumber\\
    &\ \ \ \ \left[\omega_1^2x_1^2+2\omega_1\omega_2x_1x_2+\omega_2^2x_2^2\right]\nonumber\\
    &=\omega_1\omega_2\left(x_1^2-2x_1x_2+x_2^2\right)\nonumber\\
    &\propto\frac{f(\eta_1)^2f(\eta_2)^2}{F(\eta_1)S(\eta_1)F(\eta_2)S(\eta_2)}(x_1-x_2)^2\label{eq:pre_WC}
\end{align}
where for $i=1,2$, $\eta_i=\beta_0+\beta_1x_i,
    F(\eta_i)=\int_{-\infty}^{\eta_i}f(s)ds,
    S(\eta_i)=1-F(\eta_i)$. Plug-in $\eta=\beta_0+\beta_1x$, then we have $$\det\mathfrak{M}(\xi)=\frac{f(\eta_1)^2f(\eta_2)^2}{F(\eta_1)S(\eta_1)F(\eta_2)S(\eta_2)}\left(\frac{\eta_1-\eta_2}{\beta_1}\right)^2$$
and taking the logarithm of $\det\mathfrak{M}(\xi)$ and setting the derivative w.r.t. $\eta_1$ equal to zero gives
\begin{align}
    \frac{2f'(\eta_1)}{f(\eta_1)}-\frac{f(\eta_1)}{F(\eta_1)}+\frac{f(\eta_1)}{S(\eta_1)}+\frac{2}{\eta_1-\eta_2}=0\label{eq:WC}
\end{align}
We denote the above \textbf{the key equation} and call it the \textbf{WC} equation where $WC$ stands for Wong and Cui. Now it is natural to consider, if we are given a maximizer $(\eta_1^*,\eta_2^*)$ of Formula~\ref{eq:pre_WC}, is it unique? The short answer is no unless $\eta_1^*$ and $\eta_2^*$ is symmetric around $2a$ where $a$ is a real number and we provide a lemma below.
    
    \begin{lemma}
    If $f(s)$ is symmetric, i.e., $f(a+s)=f(a-s)$ for some $a$, then for any design 
$$\xi^*=\left(\begin{matrix}
\frac{\eta_1^*-\beta_0}{\beta_1}&\frac{\eta_2^*-\beta_0}{\beta_1}\\
0.5&0.5
\end{matrix}\right)$$ 
we have
$$\det\mathfrak{M}(\xi^*)=\det\mathfrak{M}(\xi')$$
where
$$\xi'=\left(\begin{matrix}
\frac{2a-\eta_2^*-\beta_0}{\beta_1}&\frac{2a-\eta_1^*-\beta_0}{\beta_1}\\
0.5&0.5
\end{matrix}\right)$$ 
    \end{lemma}
\begin{proof}
For $i=1,2$, let $x_i^*=\frac{\eta_1^*-\beta_0}{\beta_1}$ and $x_i'=\frac{2a-\eta_2^*-\beta_0}{\beta_1}$, then
$$(x_1^*-x_2^*)^2=\left(\frac{\eta_1^*-\eta_2^*}{\beta_1}\right)^2=(x_1'-x_2')^2$$
Next, let $\eta_1'=2a-\eta_2^*$ and $\eta_2'=2a-\eta_1^*$, we have
$$f(\eta_1^*)=f(a+(\eta_1^*-a))=f(a-(\eta_1^*-a))=f(2a-\eta_1^*)=f(\eta_2')$$
and similarly, $f(\eta_2^*)=f(\eta_1')$.

Finally, we have
\begin{align*}
    F(\eta_1^*)&=F(a+(\eta_1^*-a))\\
    &=S(\eta_2')
\end{align*}
and $S(\eta_1^*)=F(\eta_2')$, $F(\eta_2^*)=S(\eta_1')$, $S(\eta_2^*)=F(\eta_1')$.
\end{proof}

\subsection{Symmetric Densities}

WLOG, we may assume that $f(s)=f(-s)$. As an example, one such $f$ is logistic density $f(s)=e^{-s}/(1+e^{-s})^2$. For a symmetric two-point design, we let $\eta_1=-\eta_2$, then

$$\det\mathfrak{M}(\xi)=\frac{f(\eta_1)^4}{F(\eta_1)^2S(\eta_1)^2}\left(\frac{2\eta_1}{\beta_1}\right)^2$$

and taking the logarithm of $\det\mathfrak{M}(\xi)$ and setting the derivative w.r.t. $\eta_1$ equal to zero gives
\begin{align}
    \frac{2f'(\eta_1)}{f(\eta_1)}-\frac{f(\eta_1)}{F(\eta_1)}+\frac{f(\eta_1)}{S(\eta_1)}+\frac{1}{\eta_1}=0
\end{align}

The resulting solutions provide the design points of the $D$-optimal and $G$-optimal designs among all $2$-point designs. To verify that it is optimal among all possible designs, we need to calculate the sensitivity function based on the Theorem~\ref{thm:equivalence}.

In the following, we apply the key equation to a few examples and verify the results using PSO . In short, we write $\eta$ for $\beta_0+\beta_1x$ and $x$ can be solved by $x=(\eta-\beta_0)/\beta_1$. 

\begin{exmp}[Logit] For this problem, $f(\eta)={\exp(\eta)}/{(1+\exp(\eta))^2}$ and $w=1/((1+\exp \eta)(1-\exp\eta))$. Plug-in all necessary elements, the key equation is
$$2-\frac{4\exp\eta}{1+\exp\eta}+\frac{2}{\eta}=0$$
Solving it numerically, we obtain $\eta_1=+1.5434$ and $\eta_2=-1.5434$. Hence, the resulting design is
\begin{align}
    \begin{cases}
    x_1=(+1.5434-\beta_0) / \beta_1\\
    x_2=(-1.5434-\beta_0) / \beta_1
    \end{cases}
\end{align}
\end{exmp}

\begin{figure}[h]
\centering
\includegraphics[width=7cm]{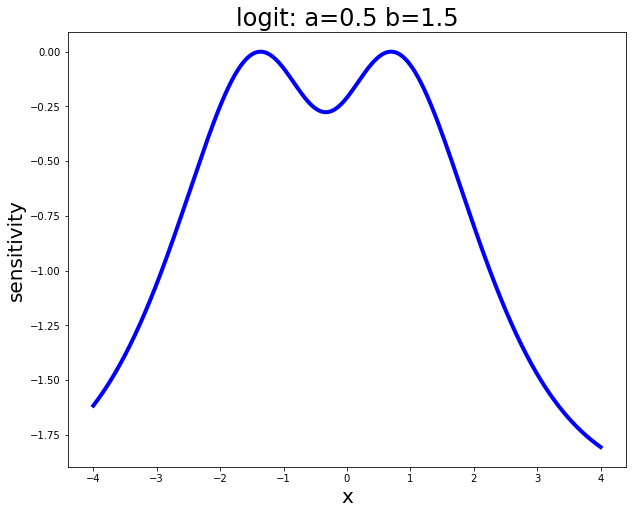}
\includegraphics[width=7cm]{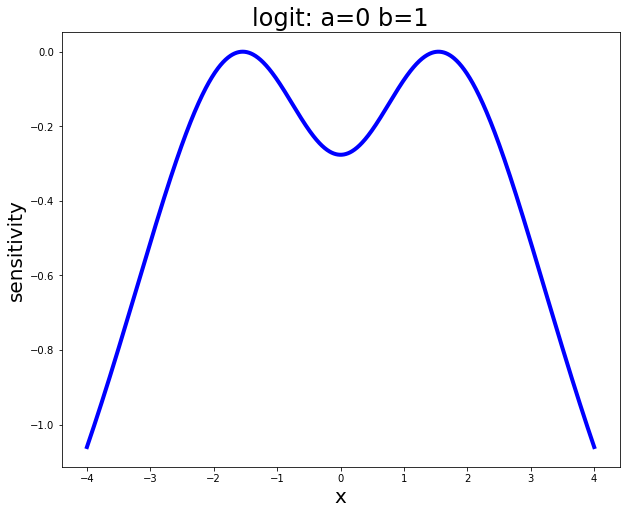}
\caption{Sensitivity functions for logit link.}
\label{fig:logit_link}
\end{figure}

The Figure \ref{fig:logit_link} demonstrates the sensitivity functions of two locally D-optimal designs with logit link and specified parameter values. 

\begin{exmp}[Probit] For this problem, $f(\eta)=\exp\left(-\frac{1}{2}\eta^2\right)/\sqrt{2\pi}$ and $\Phi(\eta)=\int_{\infty}^\eta f(s)ds$ and $w=\exp(-\eta^2)/(2\pi \Phi(\eta)(1-\Phi(\eta)))$. Plug-in all necessary elements, the key equation is
$$\frac{2}{\eta}-4\eta-2f(\eta)\left(\frac{1}{\Phi(\eta)}-\frac{1}{1-\Phi(\eta)}\right)=0$$
Solving it numerically, we obtain $\eta_1=+1.1381$ and $\eta_2=-1.1381$. Hence, the resulting design is
\begin{align}
    \begin{cases}
    x_1=(+1.1381-\beta_0) / \beta_1\\
    x_2=(-1.1381-\beta_0) / \beta_1
    \end{cases}
    \end{align}
    The two panels of Figure \ref{fig:probit_link} demonstrates the sensitivity functions of two locally D-optimal designs with probit link and specified parameter values.

\begin{figure}[h]
\centering
\includegraphics[width=7cm]{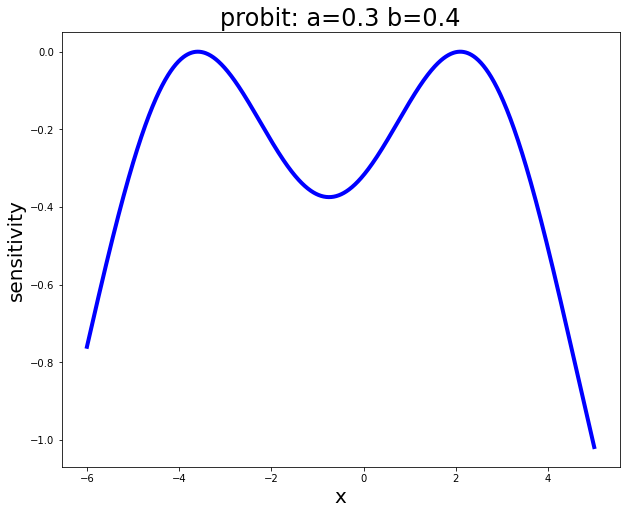}
\includegraphics[width=7cm]{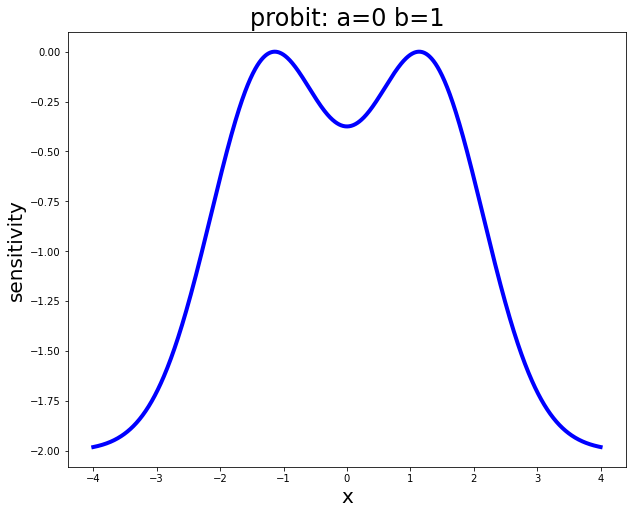}
\caption{Sensitivity functions for probit link.}
\label{fig:probit_link}
\end{figure}

\end{exmp}

\begin{exmp}[Laplace] For this problem, we have $f(\eta)=\exp(-|\eta|)/2$ and $w=1/(2\exp(|\eta|)-1)$. Plug-in all necessary elements, the key equation is
$$-4\ \text{sgn}(\eta)-\frac{2\exp(-|\eta|)}{1+\text{sgn}(\eta)(1-\exp(-|\eta|))}+\frac{2\exp(-|\eta|)}{1-\text{sgn}(\eta)(1-\exp(-|\eta|))}+\frac{2}{\eta}=0$$
Solving it numerically, we obtain $\eta_1=+0.7680$ and $\eta_2=-0.7680$. Hence, the resulting design is
\begin{align}
    \begin{cases}
    x_1=(+0.7680-\beta_0) / \beta_1\\
    x_2=(-0.7680-\beta_0) / \beta_1
    \end{cases}
    \end{align}
    
    However, the left panel of Figure \ref{fig:laplace_link} has shown that $(+0.7680, -0.7680)$ is NOT a locally D-optimal design. According to Federov's algorithm \cite{atkinson2007optimum}, it suggests that we need to add a design point at 0. This is empirically verified by Particle Swarm Optimization (PSO) using the Python package ``pyswarms'' \cite{miranda2018pyswarms}. The right panel of Figure \ref{fig:laplace_link} has shown the sensitivity function of the three point design generated by PSO.

\begin{figure}[ht]
\centering
\includegraphics[width=7cm]{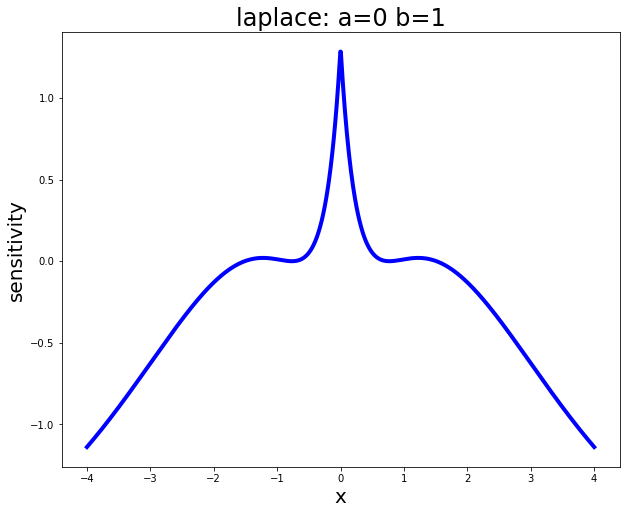}
\includegraphics[width=7cm]{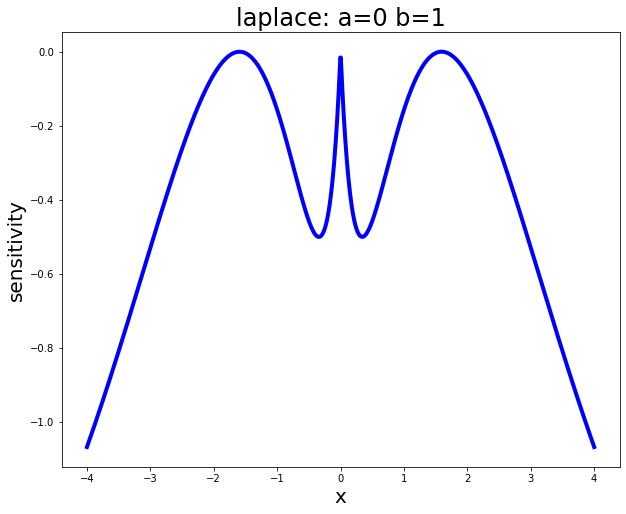}
\caption{Sensitivity functions for Laplace link.}
\label{fig:laplace_link}
\end{figure}

\end{exmp}

Now is it natural to ask that when does the two point design is indeed globally optimal, i.e., is it possible that we have a three-or-more point $D$-optimal design? Due to Caratheodory's theorem (for example, see appendix in \cite{silvey2013optimal}), the $D$-optimal design for a two parameter binary regression model has AT MOST three support points. To illustrate this idea, we add another example with an artificial regression function.

\begin{exmp}[Logit with an artificial regression function] Instead of $\eta=\beta_0+\beta_1x$, suppose now we have
$$\eta=\beta_0+\beta_1x+\frac{1}{|x|+1}$$
In this case, the WC equation boils down to
$1.5434 = \beta_0+\beta_1x+\frac{1}{|x|+1}$ and $-1.5434 = \beta_0+\beta_1x+\frac{1}{|x|+1}$. Alternatively, we solve the $D$-optimal design using PSO (50 particles with 500 iterations, hyper-parameters are $c_1= 0.5, c_2= 0.3$ and $w= 0.9$) and the results are shown in left panel of Figure~\ref{fig:logit_erratic}. Note that still we only have 2 parameters $\beta_0$ and $\beta_1$ but the $D$-optimal has $3$ support points ($x_1=0, x_2=-1.9918\text{ and }x_3=1.1242$) and $3$ is, by Caratheodory's theorem, the most number of points for a two parameter binary regression model. In this case, the three points are not equally weighted and they have weights $0.2648, 0.4289, 0.3063$ respectively.

In addition, it is not necessary that a $D$-optimal design for this artificial regression has $3$ support points. For example, if we let $\beta_0=0.1$, $\beta_1=1$ and restrict $\eta$ to $[-10,10]$, then the $D$-optimal design has only $2$ support points (right panel of Figure~\ref{fig:logit_erratic}). We have $x_1=-0.01861$ and $x_2=-10$.
\begin{figure}[h]
\centering
\includegraphics[width=7cm]{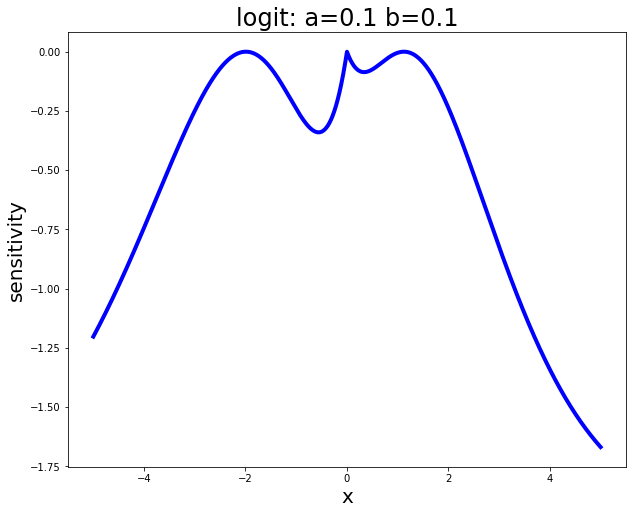}
\includegraphics[width=7cm]{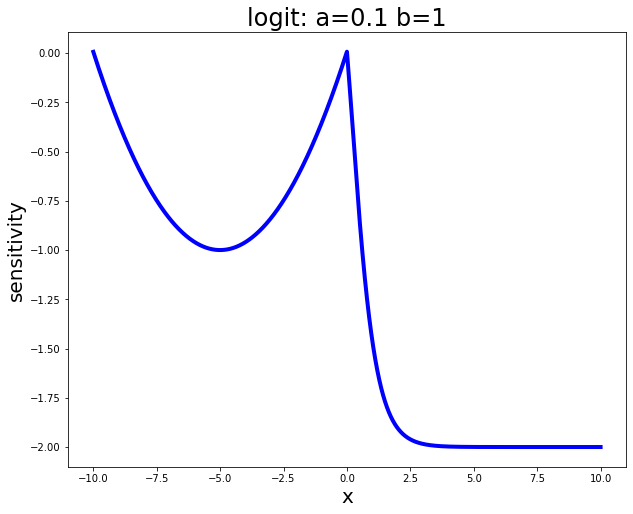}
\caption{Sensitivity functions for logit link with a erratic regression function.}
\label{fig:logit_erratic}
\end{figure}
\end{exmp}

\begin{exmp}[Student t] For this problem, $f(\eta)=\frac{\Gamma(\frac{k+1}{2})}{\sqrt{k\pi}\Gamma(\frac{k}{2})}\left(1+\frac{\eta^2}{k}\right)^{-\frac{k+1}{2}}$ and $\omega=f(\eta)^2/(F(\eta)S(\eta))$.  The Figure \ref{fig:student_link} demonstrates the sensitivity functions of two locally D-optimal designs with Student t link (degrees of freedom is 2) and specified parameter values. The left panel has design space $[-10, 10]$ and the $D$-optimal design is $x_1=-10,x_2=-0.5247$. The right panel has design space $[0, 1]$ and the $D$-optimal design is $x_1=1.7121, x_2=0$.
\begin{figure}[h]
\centering
\includegraphics[width=7cm]{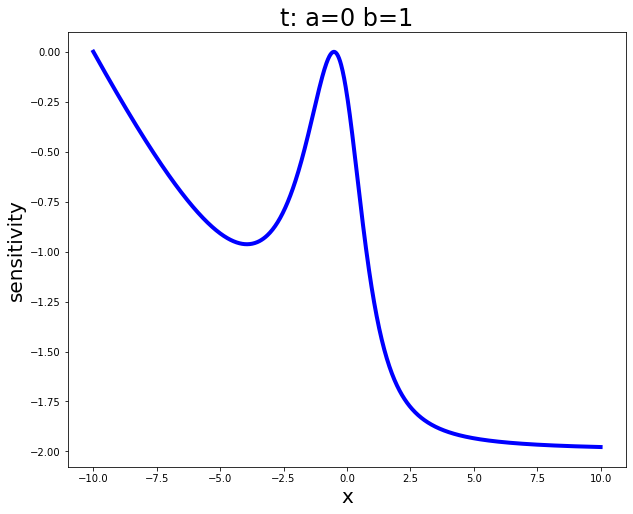}
\includegraphics[width=7cm]{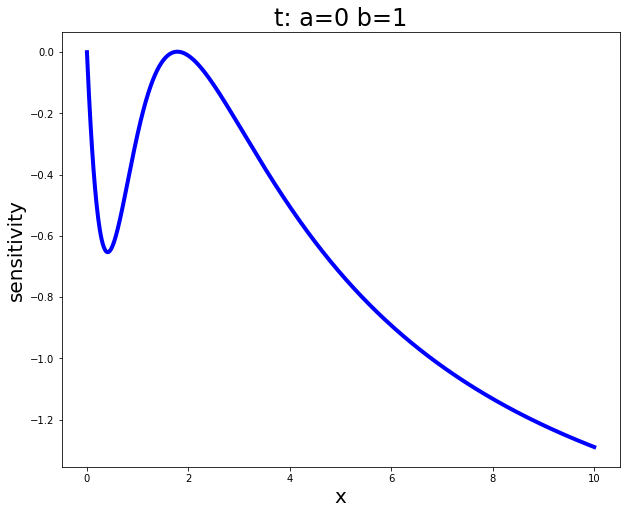}
\caption{Sensitivity functions for Student t link.}
\label{fig:student_link}
\end{figure}
\end{exmp}

\subsection{Asymmetric Densities}

If $f(\cdot)$ is asymmetric, then the resulting optimal design is not symmetric in general. In this case, we have a system of non-linear WC equations:
\begin{align*}
        \frac{2f'(\eta_1)}{f(\eta_1)}-\frac{f(\eta_1)}{F(\eta_1)}+\frac{f(\eta_1)}{S(\eta_1)}+\frac{2}{\eta_1-\eta_2}&=0\\
        \frac{2f'(\eta_2)}{f(\eta_2)}-\frac{f(\eta_2)}{F(\eta_2)}+\frac{f(\eta_2)}{S(\eta_2)}+\frac{2}{\eta_2-\eta_1}&=0
\end{align*}
Define
$$W(\eta)=\frac{2f'(\eta)}{f(\eta)}-\frac{f(\eta)}{F(\eta)}+\frac{f(\eta)}{S(\eta)}$$
and substitute it into the above system of non-linear equations:
\begin{align}
    W(\eta_1)&=-W\left(\frac{2}{W(\eta_1)}+\eta_1\right)\nonumber\\
    \eta_2&=\eta_1+\frac{2}{W(\eta_1)}
\end{align}
Hence, in practice, we solve the first equation for $\eta_1$ and then plug-in it to the second one to derive $\eta_2$.

\begin{exmp}[Cox Regression] For this problem, $f(\eta)=\exp\left(\eta-\exp(\eta)\right)$. Plug-in all necessary elements and solving numerically, we obtain $\eta_1=+0.9796$ and $\eta_2=-1.3378$.

The Figure \ref{fig:cox_link} demonstrates the sensitivity functions of two locally D-optimal designs with complementary log-log link and specified parameter values. 
\begin{figure}[h]
\centering
\includegraphics[width=7cm]{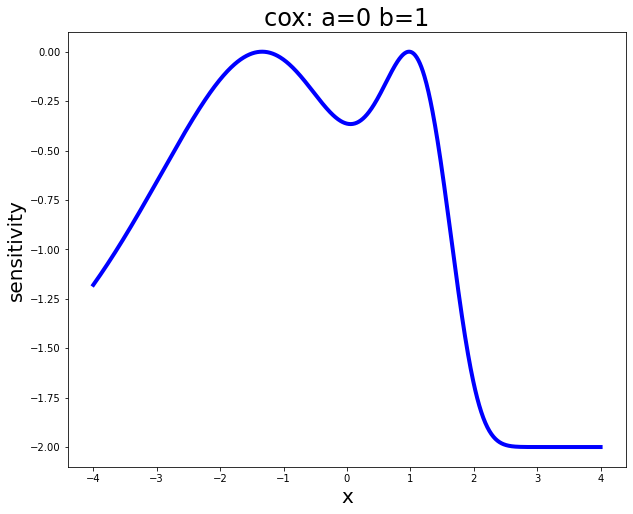}
\includegraphics[width=7cm]{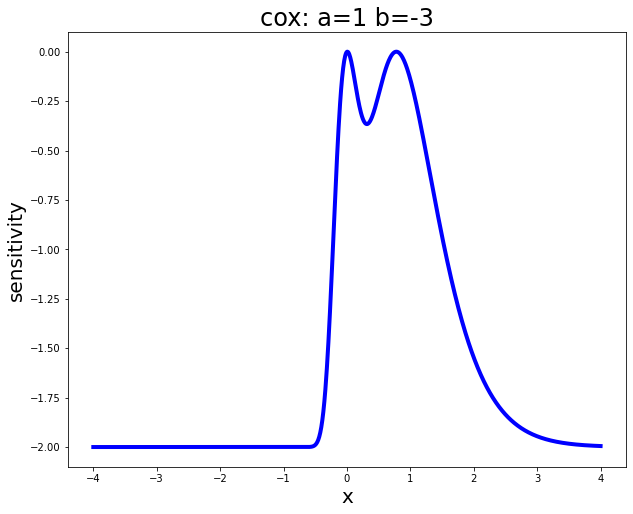}
\caption{Sensitivity functions for complementary log-log link.}
\label{fig:cox_link}
\end{figure}
\end{exmp}

\begin{exmp}[One-Hit Model] The one-hit model is also known as exponential regression in toxicology studies \cite{razzaghi2022statistical} and the density is $f(\eta)=\exp\left(-\eta\right)$ for all $\eta\ge 0$. The other terms associated with the one-hit model are $f'(\eta)=-\exp(-\eta), \pi=F(\eta)=1-\exp(-\eta) \text{ and } S(\eta)=\exp(-\eta)$. However, the non-linear WC equation is numerically unstable in this case. That is, if we let $h(\eta)=W(\eta)+W(\frac{2}{W(\eta)}+\eta)$, then
$$\lim_{\eta\downarrow0}h(\eta)=-\infty$$
To show this, we first write $W(\eta)$ and $h(\eta)$ explicitly, that is,
\begin{align*}
    W(\eta)&=-\frac{1}{1-\exp(-\eta)}\\
    h(\eta)&=-\frac{1}{1-\exp(-\eta)}-\frac{1}{1-\exp(2-2\exp(-\eta)-\eta)}
\end{align*}
If $\eta\downarrow 0$, then both terms go to negative infinity. Next, if we plug-in $\eta=1$, then $h(\eta)\approx5.2263$. By continuity of $h(\eta)$, there is at least one zero point between 0 and 1, and we plot the behavior of $h(\eta)$ near $0$ below.
\begin{figure}[h]
\centering
\includegraphics[width=12cm]{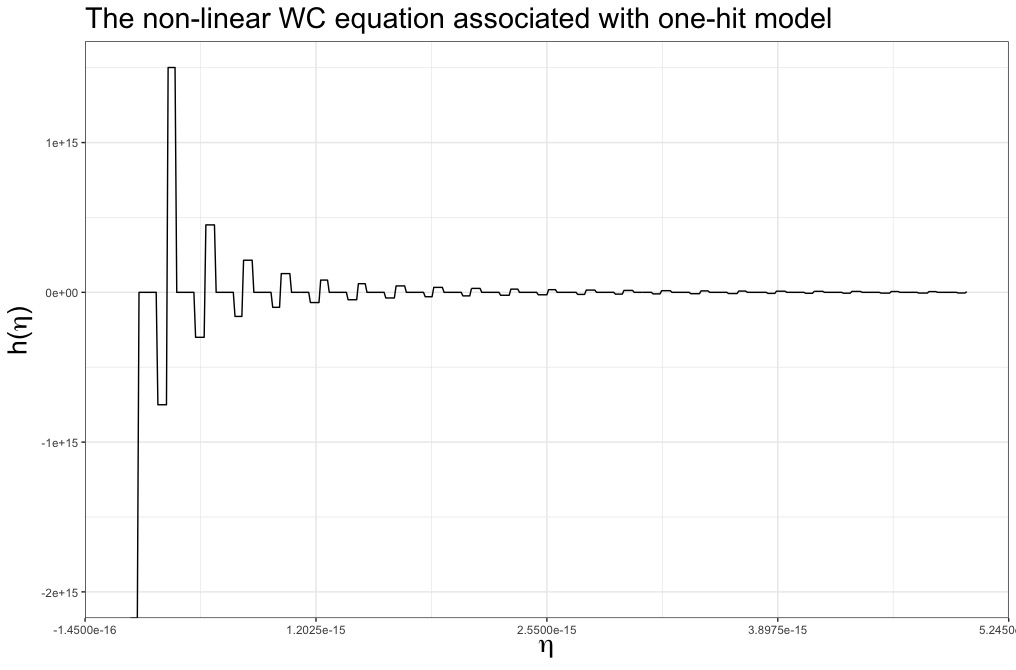}
\caption{Behavior of $h(\eta)$ near 0.}
\label{fig:exp_WC}
\end{figure}

As we can see from Figure~\ref{fig:exp_WC}, there are multiple zero points of $h(\eta)$ when $\eta$ is close to zero. Therefore, as a take home message for practitioners, it is better for us to enforce $\eta$ no less than a strictly positive value $\eta_{\text{low}}$, say $\eta_{\text{low}}=0.5$. In other words, it means that if the dose level is 0 ($x=0$), then the baseline response probability is
$$\pi=1-\exp(-\eta_{\text{low}})=1-\exp(-0.5)=0.3935$$
and the choice of $\eta_{\text{low}}$ should come from toxicologists.

The Figure \ref{fig:exp_link} demonstrates the sensitivity functions of two locally D-optimal designs with exponential link and specified parameter values. The results are generated by PSO with 50 particles and 1,000 iterations, and the hyper-parameters are $c_1= 0.5, c_2= 0.3$ and $w= 0.9$. The left panel has design space $\eta\in [0.5,1]$ and the optimal design is $x_1=0.5$, $x_2=1$. The right panel has design space $\eta\in[1,\infty]$ and the optimal design is $x_1=0$, $x_2=1.7978$. 
\begin{figure}[h]
\centering
\includegraphics[width=7cm]{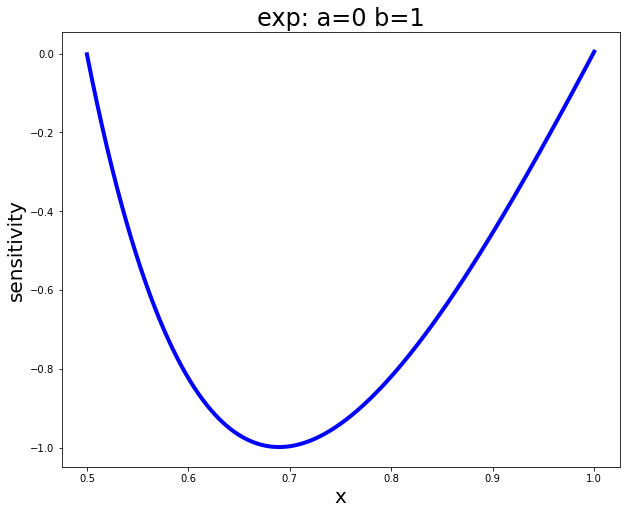}
\includegraphics[width=7cm]{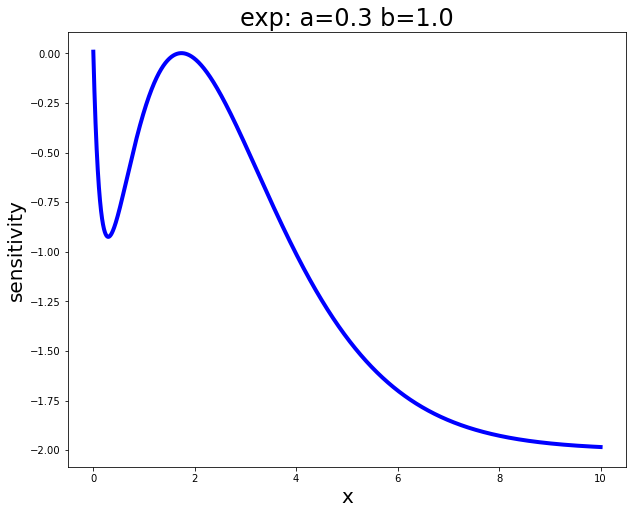}
\caption{Sensitivity functions for exponential link.}
\label{fig:exp_link}
\end{figure}

\end{exmp}

\begin{exmp}[Weibull] The Weibull distribution includes the exponential distribution as a special case. The density of Weibull distribution is $f(x)=\exp(-\beta_0-\beta_1x^\alpha)$ where $\alpha$ is a positive parameter. If $\alpha=1$, then it is the exponential density. Let $\eta=\beta_0+\beta_1x^\alpha$ then the whole framework for the one-hit model (exponential regression) can be copied almost completely. Hence, in practice, we suggest practitioners to start with a small positive $\beta_0$ to avoid the numerical issue.

The Figure \ref{fig:weibull_link} demonstrates the sensitivity functions of two locally D-optimal designs with Weibull link ($\alpha=3$) and specified parameter values. The results are generated by PSO with 50 particles and 500 iterations, and the hyper-parameters are $c_1= 0.5, c_2= 0.3$ and $w= 0.9$. The left panel has design space $\eta\in [0,3]$ and the optimal design is $x_1=0$, $x_2=0.7686$. The right panel has design space $\eta\in[0,1]$ and the optimal design is $x_1=0$, $x_2=1$. 
\begin{figure}[h]
\centering
\includegraphics[width=7cm]{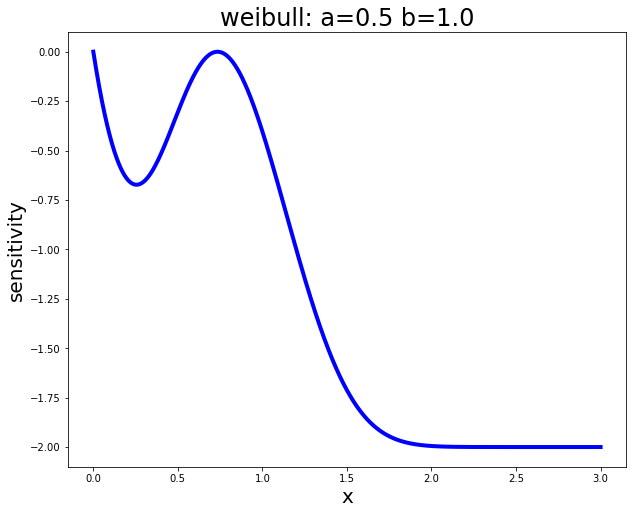}
\includegraphics[width=7cm]{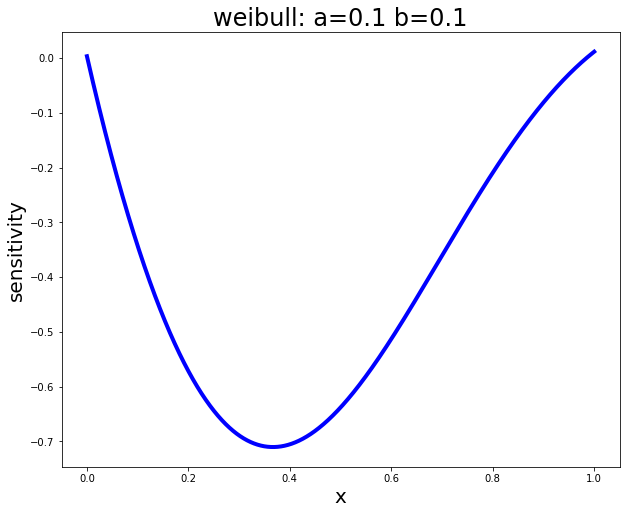}
\caption{Sensitivity functions for Weibull link.}
\label{fig:weibull_link}
\end{figure}

\end{exmp}

\section{Extension to Three-parameter Binary Regression}
Let $\pi=c+(1-c)F(\eta)$ where $c$ is the background response probability within 0 to 1 \cite{razzaghi2022statistical}. Such model is particularly useful when we have a potential baseline response probability that is strictly positive. Then the parameter now becomes $\theta^T=(c,\beta)$ where $\beta$ is a $d$-dimensional vector and in this section we always assume $d=2$. Given a single design point $x$, the Fisher information matrix of $\theta$ is ($\mathcal{L}=\log p(y|x,\pi)$)
\begin{align*}
    \mathbb{E}\left(\frac{\partial\mathcal{L}}{\partial\theta} \right)\left(\frac{\partial\mathcal{L}}{\partial\theta^T}\right)&=\mathbb{E}\left(\frac{\partial\mathcal{L}}{\partial\pi}\frac{\partial\pi}{\partial\theta} \right)\left(\frac{\partial\mathcal{L}}{\partial\pi}\frac{\partial\pi}{\partial\theta^T}\right)
\end{align*}
The term $\frac{\partial\pi}{\partial\theta}$ is a $(d+1)$-vector and we have
\begin{align*}
    \frac{\partial\mathcal{L}}{\partial\pi}&=\frac{y-\pi}{\pi(1-\pi)}=\frac{y-F(\eta)}{F(\eta)S(\eta)}\\
    \frac{\partial\pi}{\partial c}&=1-F(\eta)=S(\eta)\\
    \frac{\partial\pi}{\partial\beta}&=(1-c)f(\eta)x
\end{align*}
Plug-in them into the Fisher information matrix, we have
\begin{align*}
    \mathbb{E}\left(\frac{\partial\mathcal{L}}{\partial\theta} \right)\left(\frac{\partial\mathcal{L}}{\partial\theta^T}\right)&=\mathbb{E}\left(\frac{y-F(\eta)}{F(\eta)S(\eta)}\right)^2\left(\begin{matrix}
    S(\eta)\\(1-c)f(\eta)x
    \end{matrix}\right)\left(\begin{matrix}
    S(\eta)\\(1-c)f(\eta)x
    \end{matrix}\right)^T\\
    &=\frac{1}{F(\eta)S(\eta)}\left(\begin{matrix}
    S(\eta)^2& (1-c)S(\eta)f(\eta)x^T\\
    (1-c)f(\eta)S(\eta)x&(1-c)^2f(\eta)^2xx^T
    \end{matrix}\right)
\end{align*}
Hence, for a $k$-point design, the information matrix is
\begin{align*}
    \mathfrak{M}(\xi)=\sum_{i=1}^kp_i\left(\begin{matrix}
    \frac{S(\eta_i)}{F(\eta_i)}&\frac{f(\eta_i)}{F(\eta_i)}(1-c)x_i^T\\\\
    \frac{f(\eta_i)}{F(\eta_i)}(1-c)x_i&\frac{f(\eta_i)^2}{F(\eta_i)S(\eta_i)}(1-c)^2x_ix_i^T
    \end{matrix}\right)
\end{align*}
where $p_i$'s are weights summing to $1$. For example, suppose we are interested in $3$-point $D$-optimal design, i.e., $k=3$, then by theorem~\ref{thm:property_d_optim} we have $p_1=p_2=p_3=1/3$. The determinant of the resulting information matrix has a neat representation and we state and prove it below.
\begin{lemma}The determinant of $\mathfrak{M}(\xi)$ for $k=3$ and $p_i=1/3, i=1,2,3$ has the following representation
\begin{align}\label{eq:three_para_binary}
    \det\mathfrak{M}(\xi)\propto\left(\sum_{i=1} ^3\frac{S(\eta_i)}{F(\eta_i)}\right)^3 \det\left(\text{Var}(\widetilde{x})\right)
\end{align}
where $\widetilde{x}_i=\frac{f(\eta_i)}{S(\eta_i)}(1-c)x_i$ (a $2$-dimensional vector) and  $\text{Var}(\widetilde{x})$ is the $2\times 2$ variance-covariance matrix of $\widetilde{x}$ with respect to the tilted probability measure \cite{dabrowska2019advanced, wainwright2019high} supported at $(\widetilde{x}_1,\widetilde{x}_2,\widetilde{x}_3)$ with probability proportional to $\left(\frac{S(\eta_1)}{F(\eta_1)}, \frac{S(\eta_2)}{F(\eta_2)}, \frac{S(\eta_3)}{F(\eta_3)}\right)$. Therefore, we expression the determinant of a $3\times 3$ matrix in terms of a $2\times 2$ matrix.
\end{lemma}
\begin{proof}
By the block matrix determinant formula \cite{boyd2004convex}, we have
\begin{align*}
    \det\mathfrak{M}(\xi)\propto\left(\sum_{i=1} ^3\frac{S(\eta_i)}{F(\eta_i)}\right)^{-1}\det\left(\underbrace{\left(\sum_{i=1} ^3\frac{S(\eta_i)}{F(\eta_i)}\right)\left(\sum_{i=1}^3\frac{f(\eta_i)^2(1-c)^2}{F(\eta_i)S(\eta_i)}x_ix_i^T\right)-\left(\sum_{i=1}^3\frac{f(\eta_i)}{F(\eta_i)}(1-c)x_i\right)^{\otimes 2}}_{(\Delta)}\right)
\end{align*}
where $y^{\otimes 2}=yy^T$ and we write $\otimes 2$ for convenience. Next, we re-write $(\Delta)$ as
\begin{align*}
    (\Delta)&=\left(\sum_{i=1} ^3\frac{S(\eta_i)}{F(\eta_i)}\right)\left(\sum_{i=1}^3\frac{S(\eta_i)f(\eta_i)^2}{F(\eta_i)S(\eta_i)^2}(1-c)^2x_ix_i^T\right)-\left(\sum_{i=1}^3\frac{S(\eta_i)f(\eta_i)}{F(\eta_i)S(\eta_i)}(1-c)x_i\right)^{\otimes 2}\\
    &=\left(\sum_{i=1} ^3\frac{S(\eta_i)}{F(\eta_i)}\right)\left(\sum_{i=1}^3\frac{S(\eta_i)}{F(\eta_i)}\left(\frac{f(\eta_i)}{S(\eta_i)}(1-c)x_i\right)^{\otimes 2}\right)-\left(\sum_{i=1}^3\frac{S(\eta_i)}{F(\eta_i)}\left(\frac{f(\eta_i)}{S(\eta_i)}(1-c)x_i\right)\right)^{\otimes 2}\\
    &=\left(\sum_{i=1} ^3\frac{S(\eta_i)}{F(\eta_i)}\right)^2\left[\left(\sum_{i=1}^3\frac{S(\eta_i)/F(\eta_i)}{\sum_{j=1}^3S(\eta_j)/F(\eta_j)}\widetilde{x}^{\otimes 2}\right)-\left(\sum_{i=1}^3\frac{S(\eta_i)/F(\eta_i)}{\sum_{j=1}^3S(\eta_j)/F(\eta_j)}\widetilde{x}\right)^{\otimes 2}\right]\\
    &=\left(\sum_{i=1} ^3\frac{S(\eta_i)}{F(\eta_i)}\right)^2 \text{Var}(\widetilde{x})
\end{align*}
where $\text{Var}(\widetilde{x})$ is the variance-covariance matrix with respect to the tilted probability measure \cite{wainwright2019high} supported at $(\widetilde{x}_1,\widetilde{x}_2,\widetilde{x}_3)$ with probability $(S(\eta_1)/F(\eta_1), S(\eta_2)/F(\eta_2), S(\eta_3)/F(\eta_3))$ and $\widetilde{x}_i=\frac{f(\eta_i)}{S(\eta_i)}x_i$. Substituting the new expression of $(\Delta)$ into $\det\mathfrak{M}(\xi)$, we have the neat formula~\ref{eq:three_para_binary} for the determinant of $\mathfrak{M}(\xi)$.
\end{proof}
Unfortunately, an analytical maximizer (or a similar WC equation for formula~\ref{eq:three_para_binary}) is difficult to obtain and not useful in practice. Hence, we suggest to apply formula~\ref{eq:three_para_binary} and optimization tools such as PSO to solve for the optimal design in practice. Further, if the interest is in estimating $c$ in particular, then the corresponding $D_s$-optimality is the Schur complement \cite{boyd2004convex} of the first element of $\mathfrak{M}(\xi)$, that is, we seek a design $\xi$ that maximizes
$$\left(\sum_{i=1} ^3p_i\frac{S(\eta_i)}{F(\eta_i)}\right)-\left(\sum_{i=1}^3p_i\frac{f(\eta_i)}{F(\eta_i)}x_i^T\right)\left(\sum_{i=1}^3\frac{f(\eta_i)^2}{F(\eta_i)S(\eta_i)}x_ix_i^T\right)^{-1}\left(\sum_{i=1}^3p_ix_i\frac{f(\eta_i)}{F(\eta_i)}\right)$$
It is worth noting that $D_s$-optimality may lead to singular information matrix $\mathfrak{M}(\xi)$ and in this case, only certain linear combinations of the parameters are estimable \cite{pazman1986foundations, cui2021lecnotes, silvey2013optimal}. Further, the sensitivity function for model $\pi=c+(1-c)F(\eta)$ is different from the one provided in \ref{thm:equivalence} and by the linearity technique \cite{pazman1986foundations, silvey2013optimal}, we define
\begin{align}
    \psi(\mathbf{x},\theta)&=\mathbf{u}^T\mathfrak{M}(\xi)^{-1}\mathbf{u} - d\\
    \mathbf{u}^T&=\left(\begin{matrix}S(\eta),& (1-c)\sqrt{\omega(\mathbf{x})}\mathbf{x}\end{matrix}\right)\nonumber
\end{align}
where $\omega(\mathbf{x})$ is, again, a weight depending on the link $\eta(\mathbf{x})$. The global optimality of $\xi$ can be verified using the analogue of the equivalence theorem \ref{thm:equivalence}.

\begin{exmp} (Logit) We provide two examples regarding the logit link here. In Figure~\ref{fig:three_para_logit_link}, the left panel shows the sensitivity function when $\beta_0=1,\beta_1=0.5$ and $c=0.1$ and the constraint is $[0,1]$. The locally $D$-optimal design is $\xi^*=[0, 0.4643, 1]$ with equal weights. The right panel shows the sensitivity function when $\beta_0=0,\beta_1=1$ and $c=0.2$ and the constraint is $[-10, 10]$. The locally $D$-optimal design is $\xi^*=[-10,-1.4555, 1.6137]$ with equal weights. Interestingly, both designs include the lower bound (0 and -10) as a design point, this is probably due to the accurate estimation of the baseline parameter $c$.

\begin{figure}[h]
\centering
\includegraphics[width=7cm]{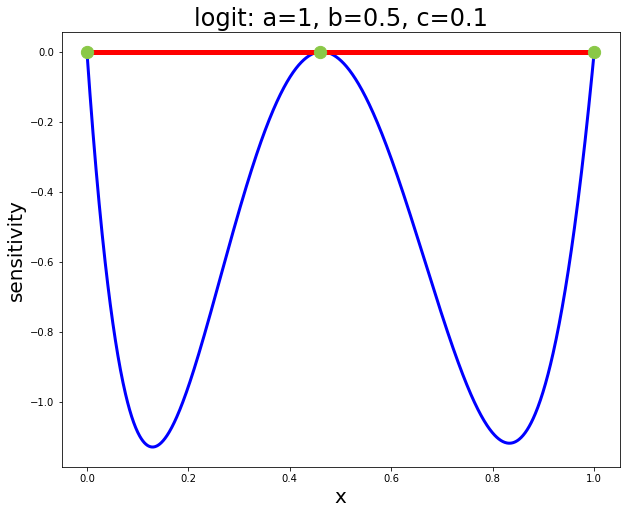}
\includegraphics[width=7cm]{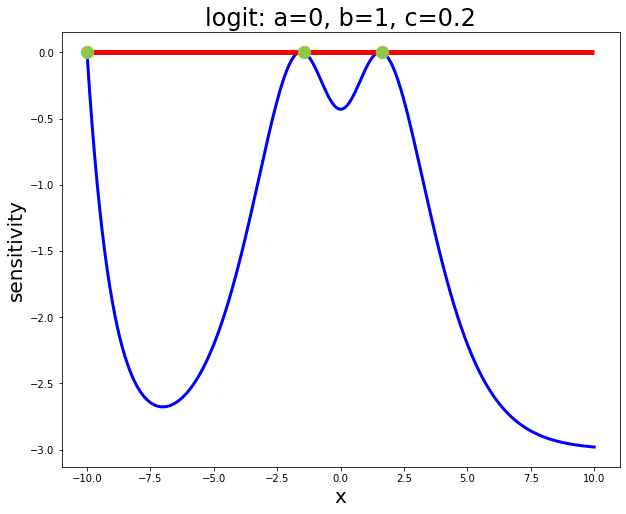}
\caption{Sensitivity functions for logit link.}
\label{fig:three_para_logit_link}
\end{figure}

\end{exmp}

\section{Applications to Toxicology Studies}

\begin{figure}[h]
\centering
\includegraphics[width=14cm]{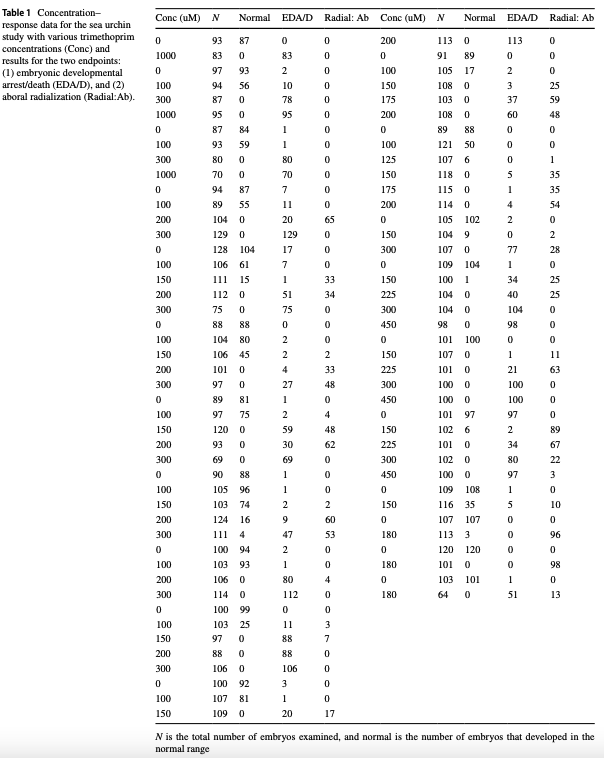}
\caption{The dataset from toxicology studies using sea urchins.}
\label{fig:data}
\end{figure}

In this section, we apply the developed WC theory to a dataset which comes from  toxicology studies using sea urchins \cite{collins2022model} (Figure~\ref{fig:data}). There are two endpoints (failure types): EDA/D and Radial:Ab and we use the second endpoint for illustration. The concentration level for the second endpoint is within 0 to 450 $\mu M$, and we re-scale it to $[0,0.45]$ by dividing 1000. We run two binary regression models using logit and complementary log-log (Cox regression) link functions respectively. The results are generated by `gtsummary` package in R \cite{sjoberg2021reproducible} and given in Table~\ref{tab:sea_urchin}. 

\begin{table}[!hbtp]
\caption{Binary regression with two different link functions using sea urchin data}
\centering
\begin{tabular}{c|ccc}
\hline \hline
\multicolumn{4}{c}{Logit link} \\
\hline
Characteristic & Estimation & 95\% CI & $p$-value \\
\hline
$\beta_0$ 	& -4.5  & (-4.7,-4.4) & < 0.001   \\
\hline
$\beta_1$  	& 20  & (19,21) & < 0.001 \\
\hline\hline
\multicolumn{4}{c}{Cox regression} \\
\hline
Characteristic & Estimation & 95\% CI & $p$-value \\
\hline
$\beta_0$ 	& -3.7  & (-3.8, -3.6) & < 0.001   \\
\hline
$\beta_1$  	& 14  & (13, 14) & < 0.001 \\
\hline \hline
\end{tabular}
\label{tab:sea_urchin}
\end{table}

\begin{figure}[!htbp]
\centering
\includegraphics[width=11cm]{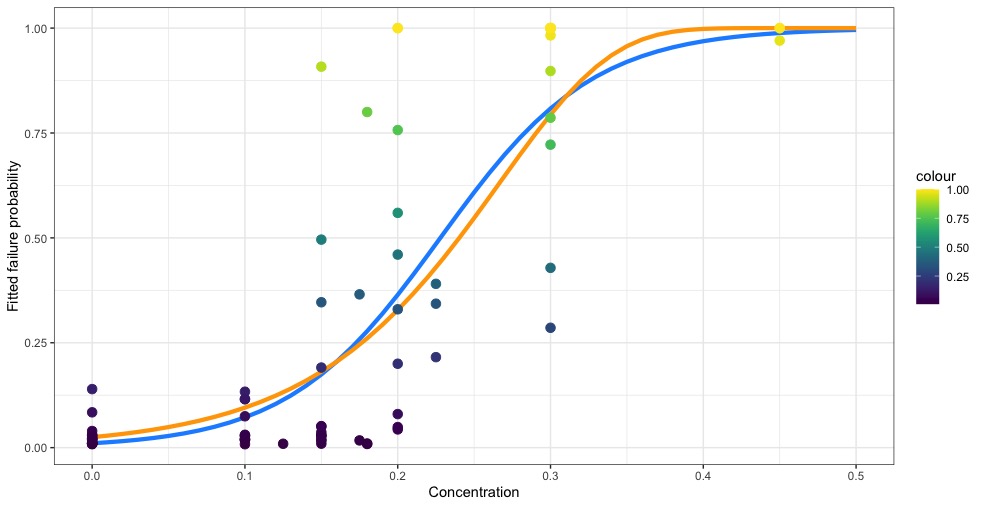}
\caption{The fitted concentration-response curves.}
\label{fig:fitted_cox}
\end{figure}

\begin{figure}[h]
\centering
\includegraphics[width=7cm]{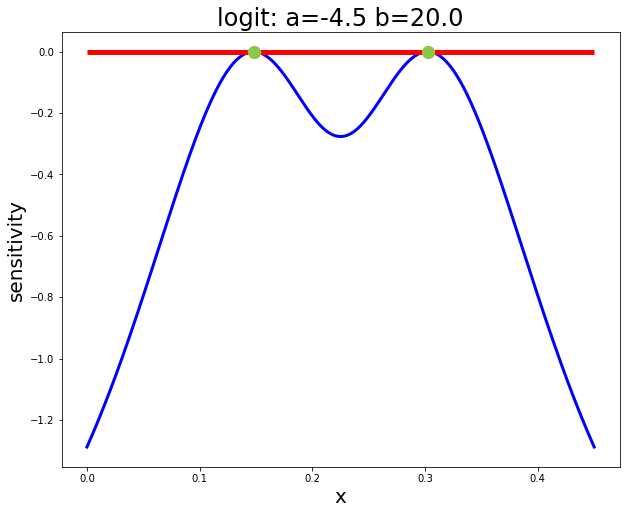}
\includegraphics[width=7cm]{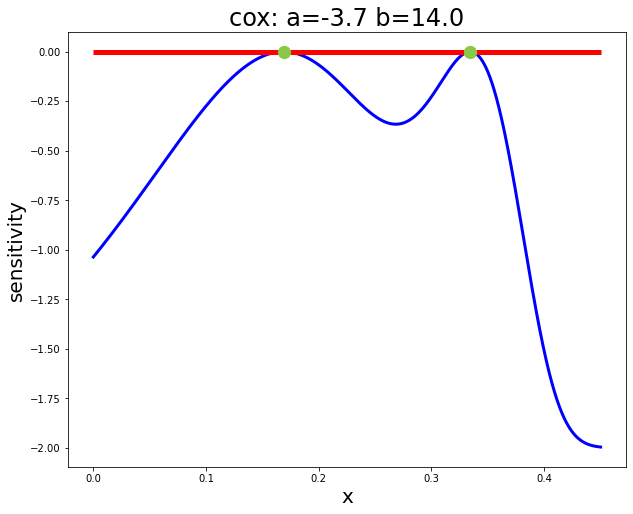}
\caption{Sensitivity functions for the sea urchin data.}
\label{fig:sea_urchin_optim}
\end{figure}

The fitted dose-response curve (in this case, concentration-response curve) is given in Figure~\ref{fig:fitted_cox}: the orange and dodgerblue curves correspond to Cox regression and logit link respectively. The dots represent the true observations from Table~\ref{fig:data} with concetration level greater than 450 removed.
Hence, by the WC equation \ref{eq:WC} for two-parameter binary regression, the $D$-optimal designs are
\begin{align*}
    \xi_{\text{logit}}&=\left(\begin{matrix}0.1478 & 0.3022\\ 0.5 & 0.5\end{matrix}\right)\\
    \xi_{\text{Cox}}&=\left(\begin{matrix}0.1687 & 0.3343\\ 0.5 & 0.5\end{matrix}\right)
\end{align*}
and the sensitivity functions are given in Figure~\ref{fig:sea_urchin_optim} (left panel: logit link; right panel: Cox regression). 
Multiplying by 1000, the resulting $D$-optimal designs at the original scale are
\begin{align}
    \xi^*_{\text{logit}}&=\left(\begin{matrix}147.8 & 302.2\\ 0.5 & 0.5\end{matrix}\right)\\
    \xi^*_{\text{Cox}}&=\left(\begin{matrix}168.7 & 334.3\\ 0.5 & 0.5\end{matrix}\right)
\end{align}
Comparing them with the original design given in \cite{collins2022model}:
\begin{align}
    \xi^*_{\text{original}}&=\left(\begin{matrix}
    0&100&125&150&175&180&200&225&300&450\\
    0.254&0.148&0.0129&0.169&0.0263&0.0338&0.128&0.0370&0.155&0.0360\end{matrix}\right)
\end{align}
we find that the $D$-optimal design reduces the number of required concentration levels significantly.

\section{Discussion}

In this paper, we have systematically discussed the $D$-optimal design in two-parameter binary regression model with various link functions. As an extension, we have provided a analytical formula of determinant for handling three-parameter binary regression. PSO, a type of metaheuristics \cite{chen2022particle}, is applied to derive $D$-optimal designs when analytical solution is not available by the WC equation. For four- and more parameter designs, the analytical formula is neither easy to derive nor useful in practice. Therefore, we suggest practitioners to use PSO to find optimal designs instead of working with the analytical solutions. We have also provided a real data example in toxicology studies illustrating the use of the developed results.

To handle more complicated situations (i.e., multi-hills of dose-response curve), we can further extend the function $\pi=F(\eta)$ to multistage \cite{carlborg1981multi}, multi-hit \cite{rai1981generalized} and dichotomous hill models \cite{gutting2016dose}. We leave these as future work and emphasis that metaheuristics can be applied to these models conveniently compared with analytical solutions.

\newpage

\newpage
\bibliographystyle{alpha}  
\bibliography{references}

\end{document}